\documentclass[eqsecnum,aps, twocolumn, superscriptaddress, 9pt]{revtex4-1}
\usepackage[textwidth=16cm,textheight=25cm]{geometry}
\usepackage[pdftex]{graphicx}
\usepackage{epstopdf}
\usepackage{hyperref}
\usepackage{amsmath}

\begin{document}

\title{ Relativistic coupled cluster calculations on hyperfine structures and electromagnetic transition amplitudes of In III}
\author{ \large Sourav Roy, \large Narendra Nath Dutta and \large Sonjoy Majumder \\
 \small {\it Department of Physics, Indian Institute of Technology-Kharagpur, \\
\small   Kharagpur-721302, India \\
 }  }

\date{\today}

\begin{abstract}
Hyperfine constants and anomalies of ground as well as few low lying
excited states of $^{113,115,117}$In III are studied with highly correlated relativistic coupled-cluster theory. The ground state hyperfine
splitting of $^{115}$In III is estimated to be 106.8 GHz.
A shift of almost 1.9 GHz of the above frequency has been calculated due to modified nuclear dipole moment. This splitting result shows its applicability as communication band and frequency standards at $10^{-11}$ sec. Correlations study of hyperfine constants indicates a few distinct features of many-body effects in the wave-functions in and near the nuclear region of this ion. Astrophysically important forbidden transition amplitudes are estimated for the first time in the literature to our knowledge. The calculated oscillator strengths of few allowed transitions are compared with recent experimental and theoretical results wherever available. 

\end{abstract}

\maketitle

\section{Introduction}

Recent trapping of doubly ionized Yb \cite{schauer} can enhance the possibility of the trapping of doubly ionized In using the same experimental mechanism. Trapped $^{113}$Cd II is known as an ideal candidate for the frequency standard and quantum computing \cite{jelenkovic,dixit}. Being in the same Ag-isoelectronic sequence, $^{113, 115, 117}$In III is further interesting from the point of view due to its expected large hyperfine splitting like $^{113}$Cd$^+$ as it is next to the latter ion in this sequence, has nuclear $g$-factor close to that of $^{113}$Cd and has large nuclear spin (I=4.5 for all these In isotopes, but I=1.5 for $^{113}$Cd). The discrepancies between experimental \cite{majumder,eck,Mann} and theoretical \cite{Chaudhuri,safronovaano} estimations of 
magnetic dipole hyperfine constants of $^{115}$In I indicate the possibility of small variation of nuclear moment obtained from Raghavan et al.\cite{raghavan}. This is, however, very important for the nuclear theory and physics where precise nuclear parameters are important, like PNC estimations \cite{Fortson,Marie,gustavsson}. The study of nuclear magnetization distribution on the indium isotopes has been an interesting topic from the point of its "puzzling giant hyperfine anomaly"\cite{niesen, persson, LUTZ}. The magnetic property can be estimated from their hyperfine anomalies (Bohr-Weisskopf effect only\cite{BW,butt}). These are calculated using accurate estimations of the hyperfine splitting of states. Study of this anomaly over different level of atomic ionizations provides the ionization effect on the nucleus. The results obtained should provide a useful calibration for nuclear theory
as well as they reduce limitations of precise measurement of 
fundamental constants, like parity and/or time violation constants,
due to the uncertainty of neutron distribution \cite{gustavsson, stroke}.

Strong resonance lines from In III are required in order to
provide the most important data for abundance analysis in various
astronomical systems as well as laboratory plasmas
\cite{TF,ZM,CR,vitas}. Explanation of the observed large discrepancies between experimental\cite{LI,DJ} and theoretical\cite{Dimitrijevic} line broadening results in the optical spectrum  \cite{LI} of In III requires precise estimations of allowed transitions. Also, the forbidden transitions are the effective decay mechanism in low density hot
plasmas where the possibility of collisional de-excitation is low
\cite{ali}. There have been many experimental and theoretical
endeavor of estimating the strengths of some ultraviolet or
visible lines of this ion over the years
\cite{ND,SK,TA,SV,GLO,cheng,martin,XB}. However, due to the large
discrepancies among the results, it is required to have
correlation exhaustive relativistic ab initio calculations. Also, we augment
the database with lifetime of few other low lying
states. Some of these are estimated for the first time in the literature
to our knowledge.

Here, we have employed highly correlated relativistic
coupled-cluster (RCC) method based on the Dirac-Coulomb-Gaunt
Hamiltonian to generate the ground and different excited states
\cite{dutta} of In III. With respect to the other well known theories,
the coupled cluster theory has the potential of taking electron
correlation in an exhaustive way \cite{dixit,Dutta}. The various
kind of many body effects like core correlation, core polarization
and pair correlation are also studied in the framework of the present RCC theory in the calculations of the hyperfine constants.

\section{Theory}

In order to obtain a correlated wave-function $\vert\Psi_v\rangle$ corresponding to a
single valence atomic state having valence electron in `$v$'th orbital,
one need to solve the corresponding energy eigen-value equation where Dirac-Coulomb-Gaunt Hamiltonian is considered\cite{dutta}.
In the coupled-cluster theory, one can write this correlated wave-function
as \cite{dixit,Dutta,Bishop,Lindgren,Sahoo},
\begin{equation}\label{1}
\vert\Psi_{v}\rangle=e^{T}\{1+S_{v}\}\vert\Phi_{v}\rangle
\end{equation}
Here, $|\Phi_{v}\rangle$ is the Dirac-Fock reference state wavefunction which is generated in the $V^{N-1}$ potential following Koopman's theorem \cite{szabo}. $T$ is the closed-shell cluster operator which takes all the single, double, and so on excitations from the core orbitals \cite{Dutta}. $S_v$ is the open-shell cluster operator which behaves as $T$ but excites atleast one electron from the valence `$v$'\cite{Dutta}.  

The general matrix element of an operator $\hat{O}$ can be
conveniently expressed with normalization as,
\begin{equation}\label{2}
O_{fi} = \frac{\langle\Psi_{f}\vert \widehat{O}\vert
\Psi_{i}\rangle}{\sqrt{ \langle \Psi_{f}\vert \Psi_{f}\rangle
\langle \Psi_{i}\vert \Psi_{i} \rangle} }
\end{equation}

The single particle reduced matrix elements of the electric dipole
($E1$), electric quadrupole ($E2$), magnetic dipole ($M1$)
transition operators and the operators associated with the
magnetic dipole ($A$) and electric quadrupole hyperfine
($B$) constants are given in Ref. \cite{sourav}.

The hyperfine anomaly due to Bohr–Weisskopf effect for any particular state is defined by the following expression \cite{Lutz,persson}
\begin{equation}\label{3}
\Delta\%=\frac{A_{1}g_{2}-A_{2}g_{1}}{A_{2}g_{1}}\times 100
\end{equation}
where, $A_1$, $A_2$ are the hyperfine constants and $g_1$, $g_2$ are the corresponding $g$-factors of the nuclei,  of the isotopes of concern.

\section{Results and Discussions}
The procedure of generating the DF orbital bases for the correlated RCC wavefunctions used in the present calculations has been discussed earlier \cite{sourav, Dutta2013}. In the present work, the percentage value in the correlation contribution (RCC result $-$ DF result) is defined with respect to the DF result. In Table~\ref{Table:I}, we represent ionisation potentials (IPs) of the ground
state and few low lying excited states in cm$^{-1}$. Our RCC
results are compared with the experimental values obtained from
the National Institute of Standards and Technology (NIST) \cite{NI}. The maximum difference among these values
occurs in the case of $4f$ $^{2}F_{5/2}$ state, which is about 0.44 $\%$.

In Table~\ref{Table:II}, we present the $E1$ transition amplitudes in the
length gauge form at the DF and the RCC levels. The correlation
contributions (Corr) are presented in the same table. The
wavelengths of these transitions which are calculated from the IPs
of NIST are quoted at the second column of this table. The correlation effect
decreases the transition amplitudes in all the cases except the
transition $5s$ $^{2}S_{1/2}$ $\rightarrow$ $ 6p$ $^{2}P_{1/2}$. We get correlation contribution is much larger than DF values for $5s$ $^{2}S_{1/2}$ $\rightarrow$ $ 6p$ $^{2}P_{1/2,3/2}$ transitions unlike to the other cases\cite{martin}. These have been also observed in velocity gauge calculations. The
two resonance transitions $5s$ $^{2}S_{1/2}$ $\rightarrow$ $5p$ $^{2}P_{1/2,3/2}$
are about -17.2 $\%$ to -17.5 $\%$ correlated and it has been estimated that most of their
correlations come from the core polarization effect. The core
polarization is also found to be the dominating mechanism in the transitions
$5p$ $^{2}P_{3/2}$ $\rightarrow$ $5d$ $^{2}D_{3/2,5/2}$ and
$5p$ $^{2}P_{1/2}$ $\rightarrow$ $5d$ $^{2}D_{3/2}$ where total correlation
contributions are about -11.8 $\%$ to -12.2 $\%$. The oscillator
strengths of $E1$ transitions calculated from the corresponding
transition amplitudes and quoted wavelengths are also presented in Table~\ref{Table:II}. These transitions are astrophysically important \cite{ZM} and fall in the visible and ultraviolet regions of
electromagnetic spectrum.

Table II also shows the discrepancies among various theoretical and experimental results. Our highly correlated {\it ab initio} calculations show excellent agreement with the  model potential \cite{JM} and RMBPT calculations \cite{SV}. Here, the experimental results are
evaluated from the corresponding lifetime measurements
\cite{TA,WA} and NIST wavelengths \cite{NI}. Considerable differences are noted
from the experimentally measured values with all the theoretical
results for the $5p$ $^{2}P_{1/2}\rightarrow 5d$ $^{2}D_{3/2}$ and
$5p$ $^{2}P_{3/2}\rightarrow 6s$ $^{2}S_{1/2}$ transitions. Therefore,
more precise experiments are desirable for these cases. Also recent experiment \cite{LI} claims accurate estimations of absorption coefficients of 298.28, 300.808 and 524.877 nm transition lines where our calculated amplitudes can be used.

Though electromagnetically forbidden transitions do not contribute
significantly to the lifetimes of the excited states here, they
are important in different areas of physics \cite{sourav}.
There has been calculation of magnetic dipole ($M1$) transition
rate between the fine-structure states of $4f$ $^{2}F$ using
multi-configuration Dirac-Fock (MCDF) method \cite{XB}. Our calculated 
fine-structure splitting (-23 cm$^{-1}$) of this term  is
much closure to the central experimental value (8 cm$^{-1}$)\cite{NI}
compare to the MCDF calculation (-71 cm$^{-1}$). The latter calculation 
estimated comparatively large correlation contribution as the DF
value is -24 cm$^{-1}$. There is a discrepancy between the MCDF and 
our calculations of transition amplitude between
these fine structure states. where former is evaluated from their calculated transition probability and wavelength. Most probably, the discrepancy 
can be avoided with proper choice of initial and final states and use of NIST wavelength\cite{NI}.

\begin{figure*}
\centering
\includegraphics[scale=0.5]{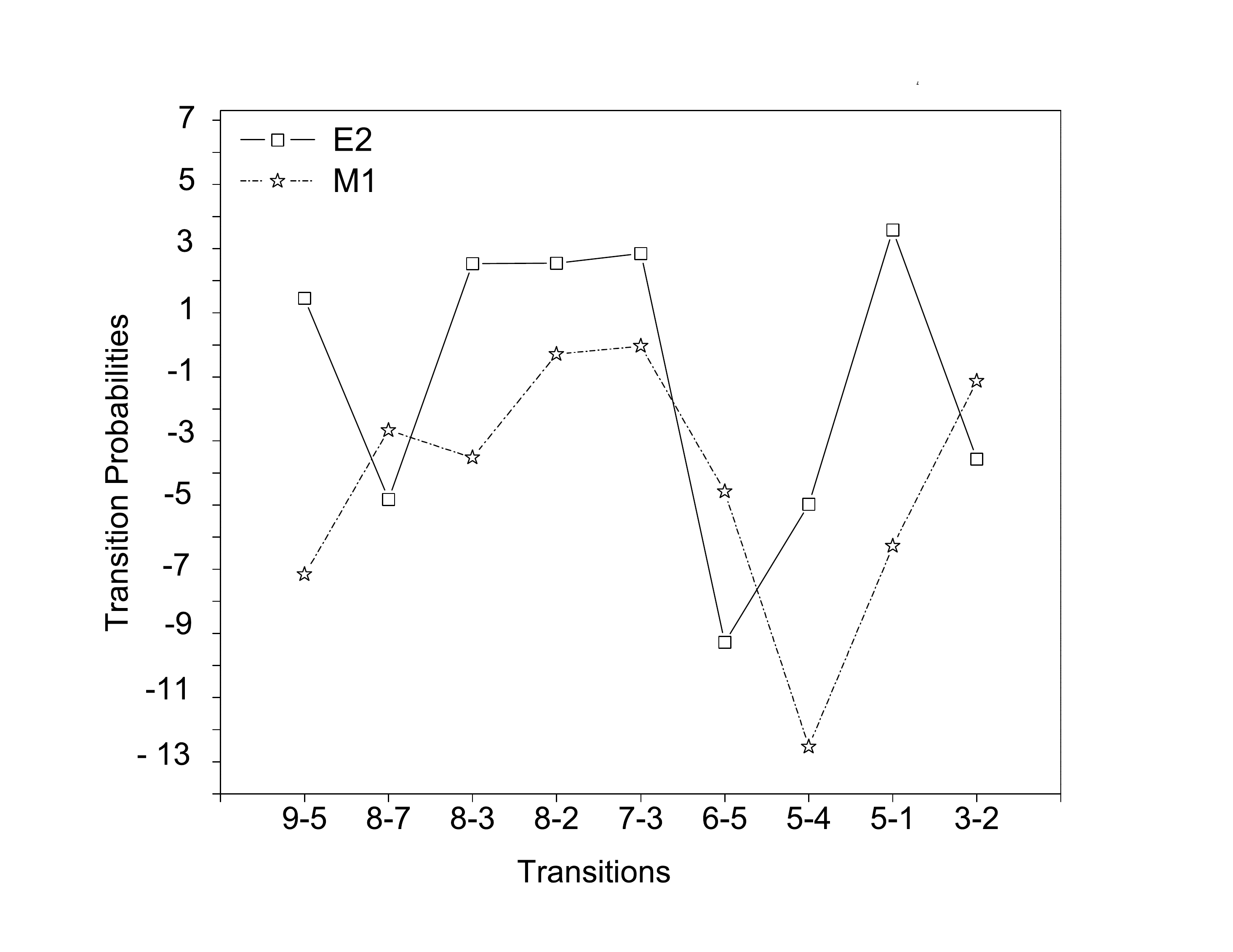}
\caption{Comparison between E2 and M1 transition probabilities ($sec^{-1}$) in the $log_{10}$ scale.\\
The transitions between levels are presented along x- axis and the transition probabilities are presented along y- axis, the levels are identified as 1: $5s$ $^{2}S_{1/2}$ , 2: $5p$ $^{2}P_{1/2}$, 3: $5p$ $^{2}P_{3/2}$, 
4: $6s$ $^{2}S_{1/2}$, 5: $5d$ $^{2}D_{3/2}$, 6: $5d$ $^{2}D_{5/2}$, 7: $6p$ $^{2}P_{1/2}$, 8: $6p$ $^{2}P_{3/2}$, 9: $7s$ $^{2}S_{1/2}$}
\label{Fig:1}
\end{figure*}

The $E2$ and $M1$ transition amplitudes along with their
corresponding NIST wavelengths ($\lambda$-values) are presented in the
Table~\ref{Table:III}. The correlation contributions (corr)
to all the $E2$ transitions reduce the corresponding DF-values and
vary from -4.9 $\%$ to -9.3 $\%$. The $5s$ $^{2}S_{1/2} \rightarrow 5d$ $^{2}D_{3/2,5/2}$ transitions are maximally correlated by -9.2 $\%$ to -9.3 $\%$ with respect to all others E2 transitions presented here. It has been observed (see Fig.~\ref{Fig:1})that M1 transition probabilities are stronger than E2 probabilities for transition among fine
structure states.

In Table~\ref{Table:IV}, we compare our calculated life times for some low
lying states with the other theoretically calculated and
experimentally measured values. Lifetimes of few states are presented
in this table for the first time in the literature to our
knowledge. The experimental wavelengths from the NIST are used in our calculations.
The beam foil experiment of Andersen et al.\cite{TA} and the relativistic Hartree Fock
(RHF) calculation \cite{cheng} of Cheng et at. underestimate the
lifetime of the $5p$ $^{2}P_{1/2}$ state. Our estimated life times are in
good agreement with the measured values of Ansbacher et al.
\cite{WA} and calculated results as obtained using the RMBPT
method \cite{SV}, except in the case of $5d$ $^{2}D_{3/2}$ state. The
lifetime of the $5d$ $^{2}D_{3/2}$ state measured by the beam foil
experiment \cite{WA} and calculated by the RMBPT method are $0.75\pm
0.06$  and $0.67$ nanosecond, respectively. These lifetimes are
based only on the transition $5d$ $^{2}D_{3/2} \rightarrow
5p$ $^{2}P_{1/2}$. Whereas, considering both the channels of
emissions $5d$ $^{2}D_{3/2} \rightarrow 5p$ $^{2}P_{1/2}$ and
$5d$ $^{2}D_{3/2} \rightarrow 5p$ $^{2}P_{3/2}$, our RCC calculations
yields this lifetime $0.53 \times 10^{-9}$ sec. Similar arguments
also hold in the comparison of the lifetimes of the
$4f$ $^{2}F_{5/2}$ state, where our calculations considered all the
channels of emissions compare to the other results which account
emission through only the dominating channel.
However, for $5d$ $^{2}D_{5/2}$ state there is only one dominating channel $5d$ $^{2}D_{5/2} \rightarrow 5p$ $^{2}P_{3/2}$ and we, therefore, find good agreement with experiment\cite{WA}. Anderson et al. also overestimates the lifetime of this state. The corrections due to the
$E2$ and/or $M1$ transitions in the calculations of all the lifetime values come at or
beyond the four decimal place in the unit of nanosecond. This may be important for ultra-fast spectroscopy\cite{Sancho}.

\begin{figure*}
\centering
\includegraphics[scale=0.5]{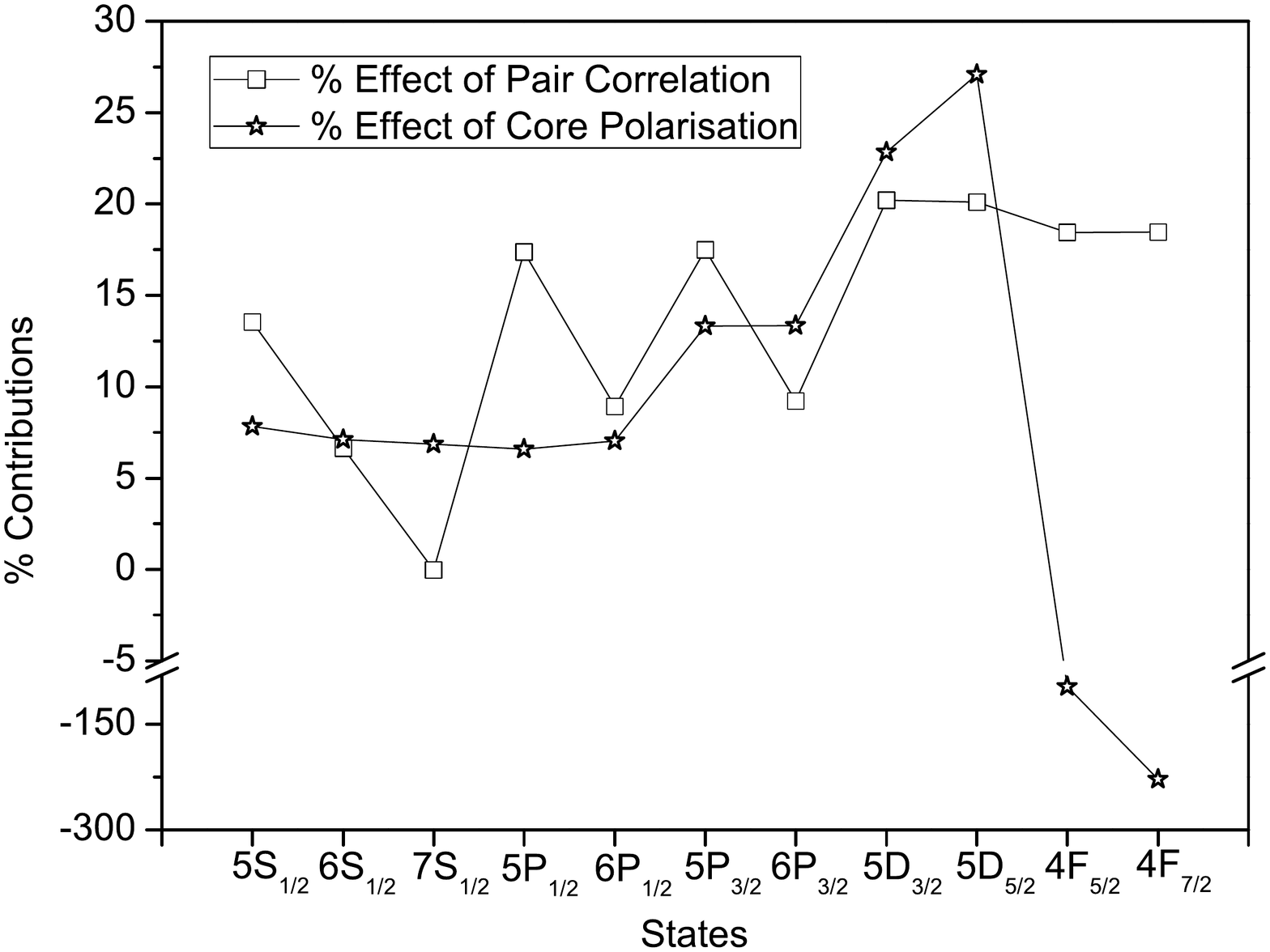}
\caption{Percentage contributions of the core polarization and pair correlation to the $A$ constants.} 
\label{Fig:2}
\end{figure*}

The accurate estimation of the large hyperfine splitting of the ground
state of $^{115}$In III is one of the most important objectives of this work. The magnetic dipole and the electric quadrupole moments of the stable $^{115}$In
isotope are considered $5.5408(2)$ nuclear magnetons and $+0.810$ barns, respectively from Ref.
\cite{raghavan} with nuclear spin-parity $9/2^{+}$. In
Table~\ref{Table:V}, the hyperfine $A$-constants of the ground and few excited
states are presented at the DF and RCC levels along with the
contributions of different correlation terms. In this table,
$\overline{d}=e^{T^{\dag}}de^{T}$ represents the sum of DF ($d$)
and core correlation contribution. According to the RCC
theory, the lowest order pair correlation and core polarization effects arise
from the terms $\overline{d}S_{1}$+c (indicated by $\overline{d}S_{1}$ in the table) and $\overline{d}S_{2}$+c (indicated by $\overline{d}S_{2}$ in the table),
respectively, where c stands for the conjugate term \cite{dutta}.
These effects are presented in Table~\ref{Table:V} along with some significantly
contributing terms like $S_{1}^{\dag}\overline {d}S_{1}$,
$S_{1}^{\dag}\overline{d}S_{2}$+c (indicated by $S_{1}^{\dag}\overline{d}S_{2}$ in the table) and $S_{2}^{\dag}\overline{d}S_{2}$. Here, 'Norm' represents the
normalization correction\cite{gopal}. The largest
correlation contributions come from the pair correlation terms for
the $5s$ $^{2}S_{1/2}$, $5p$ $^{2}P_{1/2,3/2}$ and $6p$ $^{2}P_{1/2}$
states. Whereas, the core polarization contributes the most to the
$6s$ $^{2}S_{1/2}$, $7s$ $^{2}S_{1/2}$, $5d$ $^{2}D_{3/2,5/2}$,
$6p$ $^{2}P_{3/2}$ and $4f$ $^{2}F_{5/2,7/2}$ states. These are
graphically presented in Fig.~\ref{Fig:2}. It is also clear from the Fig.~\ref{Fig:2} that the pair correlation contributions are decreasing to the outer orbitals along the same relativistic symmetry, which is expected\cite{Owusu}. The percentage of pair correlation contributes almost identically for the fine-structure states of any term has also been observed.
 $4f$ $^{2}F_{5/2}$ and $4f$ $^{2}F_{7/2}$ states show large percentage of negative core
polarization contributions. These large negative contributions
dominantly arise from the exchange part of these correlation terms.
 The $A$-constants of the low-lying bound $ns$ $^{2}S_{1/2}$
states fall in the GHz range and their high values are expected due
to the large overlap of their wave functions in the nuclear vicinity.
The total correlation contribution to the $A$-constant of the
ground state is around 20.9 $\%$. The estimated hyperfine $A$
constants of all these states are presented within an approximate theoretical
uncertainty of $\pm$2 $\%$.

\begin{figure*}
\centering
\includegraphics[scale=0.5]{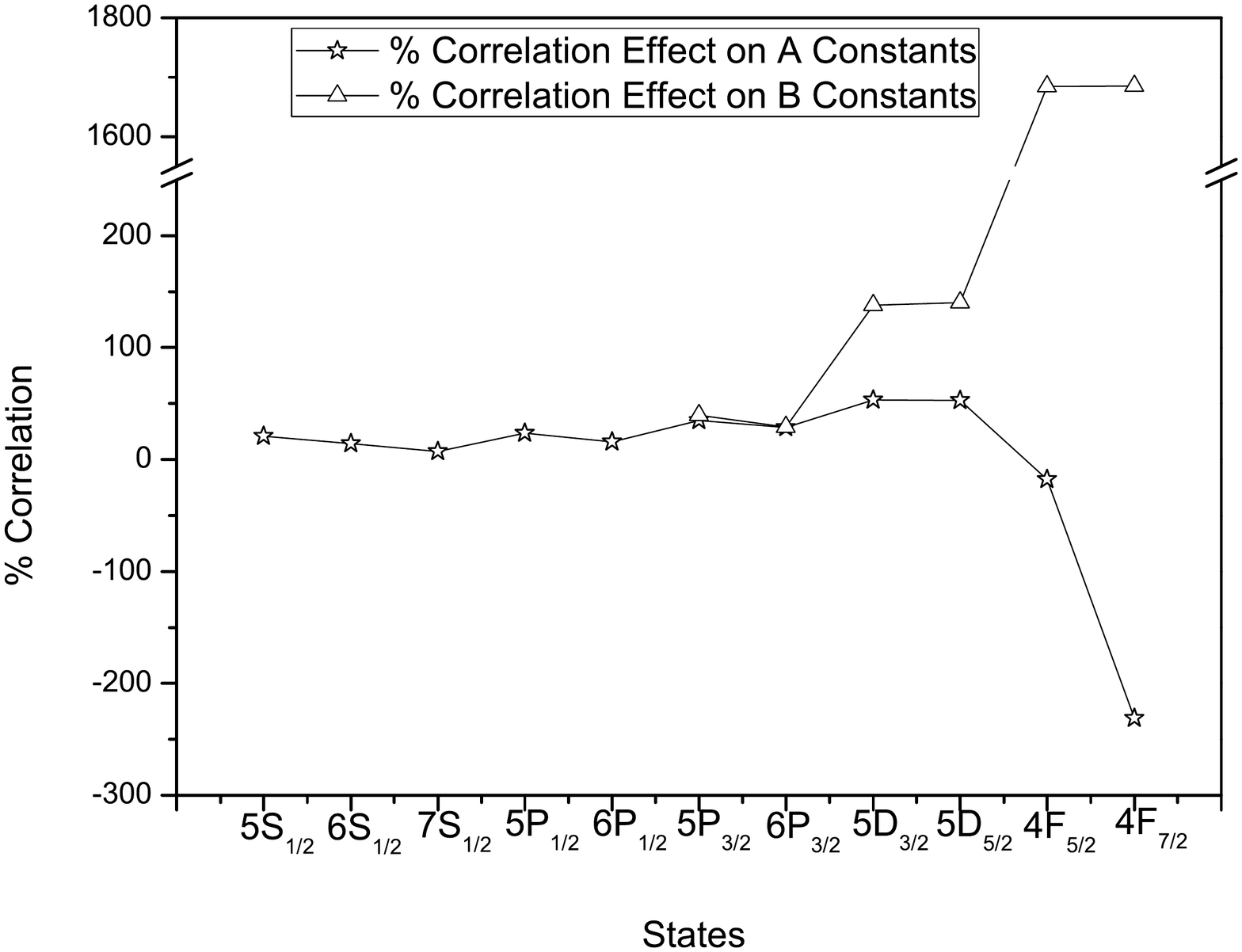}
\caption{Percentage values of correlation to the hyperfine $A$ and
$B$ constants.} 
\label{Fig:3}
\end{figure*}

\begin{figure*}

\centering
\includegraphics[scale=0.5]{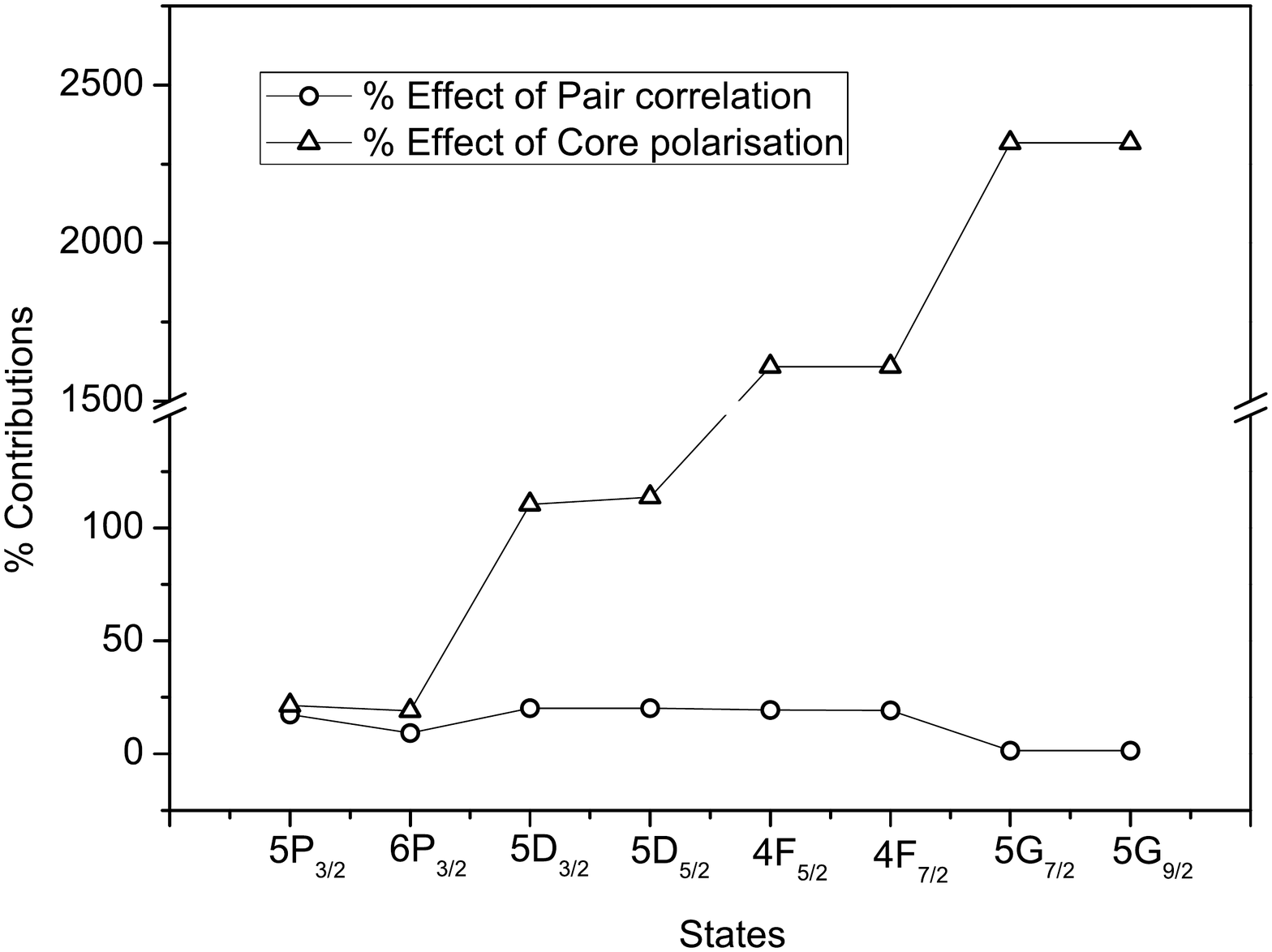}
\caption{Percentage contributions of the core polarization and pair correlation to the $B$ constants.} 
\label{Fig:4}
\end{figure*}

Our calculated hyperfine $B$-constants for the low lying states are presented in Table~\ref{Table:VI} along with the different correlation contributing many body terms. In this table, the labelling of the different terms are done identically as those are done in Table~\ref{Table:V}. The percentage values of the total correlation contributions to these $A$ and $B$ constants are plotted in Fig.~\ref{Fig:3} to get an idea about their relative responses. The correlation to the $4f$ $^{2}F$-states shows an opposite trends between these two constants. This may be a consequence of difference of the behaviour of the wavefunctions in two different radial regions of nuclear proximity. Here, one can see that
the core polarization contributes strongly to all the cases with
respect to the other correlation terms. This is clear from Table~\ref{Table:VI} and Fig.~\ref{Fig:4}. Even, the RCC values of the $5d$ $^{2}D_{3/2,5/2}$
states become more than twice of their corresponding DF values due
to large core polarization effects. The correlation contributions to the hyperfine $B$ constants of the
$4f$ $^{2}F_{5/2,7/2}$ and $5g$ $^{2}G_{7/2,9/2}$ states change
abnormally from the DF to the RCC levels. These abnormal changes are also guided by the core polarization effects. One can also find from Fig.~\ref{Fig:4} that the percentage contribution of the core polarization is almost same to the fine-structure states of a term.

The hyperfine splitting of the ground as well as few low lying excited
states are presented in Table~\ref{Table:VII}. The percentage contributions from the $B$-constants to the splitting values are presented at the last column of this table. The comparison of theoretically estimated \cite{Chaudhuri} and experimentally measured \cite{majumder} hyperfine constants of
different states shows nuclear magnetic moment may be
5.4422 $\mu_B$ for $^{115}$In I. This value varies from the standard value as obtained from Raghavan et al. by 5.5408 $\mu_B$ \cite{raghavan}. This changes the ground state splitting by 0.0634 $cm^{-1}$, which is substantial in terms of accuracy we are looking for.

Study of Nuclear magnetization distribution (NMD) of any atomic system provides informations about nuclear wave functions, which is very important for the PNC calculations\cite{stroke}. It is difficult to measure the nuclear magnetization distribution (NMD) experimentally\cite{stroke,büttgenbach}, rather it can be estimated from accurate value of hyperfine splitting using Bohr-Weisskoff formalism \cite{persson}. Though mean contribution of this effect appears only for $S_{1/2}$ and $P_{1/2}$ states, but other states, like $P_{3/2}$, are effected due to $e^{-}-e^{-}$ interaction. Using Eq(2.3),  we have calculated hyperfine anomaly of In III isotopes between 113 and 115 ($^{113}\Delta ^{115}(\%)$) and 115 and 117 ($^{115}\Delta^{117}(\%)$). We present these results in Table~\ref{Table:VIII}. Like neutral Indium, we observe considerable effect of finite nucleus in these parameters for In III\cite{Lutz}. However, we do not see significant changes in the parameter between $P_{1/2}$ and $P_{3/2}$-states as was observed in  neutral system\cite{persson,Lutz}.

\begin{table}[t]
\centering \caption{IPs of ground and low-lying excited states in
cm$^{-1}$}
\begin{tabular}{c}
\begin{tabular}{l c c }
\hline\hline
State & RCC & NIST \\
\hline
\\ 
$5s$ $^{2}S_{\frac{1}{2}} $& 226445.34 & 226191.3 \\
$5p$ $^{2}P_{\frac{1}{2}} $& 168645.05 & 169010.2 \\
$5p$ $^{2}P_{\frac{3}{2}} $& 164208.93 & 164668.1 \\
$6s$ $^{2}S_{\frac{1}{2}} $&  99353.88 &  99317.1 \\
$5d$ $^{2}D_{\frac{3}{2}} $&  97713.82 &  97738.5 \\
$5d$ $^{2}D_{\frac{5}{2}} $&  97288.94 &  97448.8 \\
$6p$ $^{2}P_{\frac{1}{2}} $&  81409.35 &  81607.7 \\
$6p$ $^{2}P_{\frac{3}{2}} $&  80011.96 &  80268.7 \\
$4f$ $^{2}F_{\frac{5}{2}} $&  63937.80 &  64222.5 \\
$4f$ $^{2}F_{\frac{7}{2}} $&  63960.98 &  64214.5 \\
$7s$ $^{2}S_{\frac{1}{2}} $&  56699.72 &  56761.5 \\
$5g$ $^{2}G_{\frac{7}{2}} $&  39497.11 &  39669.3 \\
$5g$ $^{2}G_{\frac{9}{2}} $&  39497.11 &  39669.3 \\
\hline\hline
\end{tabular}
\end{tabular}
\label{Table:I}
\end{table}

\begin{table*}
\centering \caption{Calculated E1 transition amplitudes (in a.u.)
and oscillator strengths. The corresponding wavelengths
($\lambda$) are presented in \AA . The oscillator strengths
calculated by other theory and experimental measurements (Expt)
are also reported for comparison with our relativistic
coupled-cluster (RCC) results.}
\begin{tabular}{l}
\begin{tabular}{rrrrrrrr}
\\
\hline
 & & \multicolumn{3}{|c|}{Transition amplitudes} & \multicolumn{3}{|c|}{Oscillator Strengths} \\
     Transitions & $\lambda$ & DF & Corr & RCC & RCC & Other theory & Expt  \\
\hline
\\
$5s$ $^{2}S_{1/2}\rightarrow 5p$ $^{2}P_{1/2}$ & 1748.83 & 2.0868 & -0.3656 &  1.7212 & 0.2600 &0.2519$^{a}$,0.260$^{b}$   & 0.27$^{c}$,0.2796$^{d}$\\
                                               &         &        &         &         &        & 0.260$^{e}$,0.1963$^{f}$  &                        \\
                                               &         &        &         &         &        & 0.2486$^{g}$                      
              &                        \\
                $\rightarrow 5p$ $^{2}P_{3/2}$ & 1625.40 & 2.9512 & -0.5070 &  2.4442 & 0.5647 & 0.5478$^{a}$, 0.567$^{b}$   & 0.60$^{c}$,0.5279$^{d}$\\
                                               &         &        &         &         &        & 0.278$^{e}$,0.4248$^{f}$  &                        \\
                                               &         &        &         &         &        & 0.5400$^{g}$    
              &                        \\
                $\rightarrow 6p$ $^{2}P_{1/2}$ &  691.64 & 0.0324 &  0.1130 &  0.1454 & 0.0047 &0.0003$^f$        
              &                        \\
                $\rightarrow 6p$ $^{2}P_{3/2}$ &  685.30 & 0.0403 & -0.1653 & -0.1250 & 0.0035 &0.0007$^f$ 
              &                        \\
$5p$ $^{2}P_{1/2}\rightarrow 6s$ $^{2}S_{1/2}$ & 1434.86 & 1.2795 & -0.0473 &  1.2322 & 0.1598 &0.161$^{b}$
              &                        \\
                $\rightarrow 5d$ $^{2}D_{3/2}$ & 1403.08 & 3.3519 & -0.4100 &  2.9419 & 0.9323 & 0.9113$^{a}$,0.900$^{b}$   & 0.7870$^{d}$           \\
                $\rightarrow 7s$ $^{2}S_{1/2}$ &  890.88 & 0.3917 & -0.0031 &  0.3886 & 0.0257 &
              &                        \\
$5p$ $^{2}P_{3/2}\rightarrow 6s$ $^{2}S_{1/2}$ & 1530.20 & 1.9776 & -0.0774 &  1.9002 & 0.1778 & 0.179 $^{b}$      
              & 0.2506$^{d}$           \\
                $\rightarrow 5d$ $^{2}D_{3/2}$ & 1494.11 & 1.5574 & -0.1837 &  1.3737 & 0.0954 &0.0932$^{a}$,0.092$^{b}$   &                        \\
                $\rightarrow 5d$ $^{2}D_{5/2}$ & 1487.67 & 4.6559 & -0.5481 &  4.1078 & 0.8575 &0.8387$^{a}$,0.831$^{b}$   & 0.8585$^{d}$           \\
                $\rightarrow 7s$ $^{2}S_{1/2}$ &  926.73 & 0.5786 & -0.0095 &  0.5691 & 0.0265 &             
              &                        \\
$6s$ $^{2}S_{1/2}\rightarrow 6p$ $^{2}P_{1/2}$ & 5646.72 & 4.3068 & -0.2107 &  4.0961 & 0.4573 &0.3708$^{f}$ 
              &                        \\
                $\rightarrow 6p$ $^{2}P_{3/2}$ & 5249.79 & 6.0411 & -0.2907 &  5.7504 & 0.9714 &0.7884$^{f}$  
              &                        \\
$5d$ $^{2}D_{3/2}\rightarrow 6p$ $^{2}P_{1/2}$ & 6199.32 & 4.1135 & -0.1120 &  4.0015 & 0.1983 &          
              &                        \\
                $\rightarrow 6p$ $^{2}P_{3/2}$ & 5724.16 & 1.7862 & -0.0453 &  1.7409 & 0.0407 &           
              &                        \\
                $\rightarrow 4f$ $^{2}F_{5/2}$ & 2983.65 & 7.0372 & -0.4556 &  6.5816 & 1.1110 & 1.0915$^{a}$
              & 1.1771$^{d}$           \\
$5d$ $^{2}D_{5/2}\rightarrow 6p$ $^{2}P_{3/2}$ & 5820.69 & 5.4252 & -0.1366 &  5.2885 & 0.2446 &        
              &                        \\
                $\rightarrow 4f$ $^{2}F_{5/2}$ & 3009.66 & 1.8915 & -0.1213 &  1.7702 & 0.0529 & 0.0520$^{a}$
              &                        \\
                $\rightarrow 4f$ $^{2}F_{7/2}$ & 3008.94 & 8.4595 & -0.5422 &  7.9173 & 1.0711 & 1.0394$^{a}$
              & 1.0522$^{d}$           \\
$6p$ $^{2}P_{1/2}\rightarrow 7s$ $^{2}S_{1/2}$ & 4024.76 & 2.7944 & -0.0806 &  2.7138 & 0.2764 &
              &                        \\
$6p$ $^{2}P_{3/2}\rightarrow 7s$ $^{2}S_{1/2}$ & 4254.02 & 4.2633 & -0.1228 &  4.1405 & 0.3035 & 
              &                        \\
\\
\hline
\\
\end{tabular}
\\

$a,g\rightarrow$ Third-order Relativistic Many-Body Perturbation Theory (RMBPT): \cite{SV,chou};\\
$b\rightarrow$ Core Polarized augmented Dirac-Fock method (DF+CP): \cite{JM};\\
$c\rightarrow$ Beam Foil technique : \cite{TA};\\
$d\rightarrow$ Beam Foil technique: \cite{WA};\\
$e\rightarrow$ Configuration Interaction calculation(CI): \cite{GLO};\\
$f\rightarrow$ Relativistic Quantum Defect Orbital method(RQDO): \cite{martin};
\end{tabular}
\label{Table:II}
\end{table*}

\begin{table*}
\centering
\setlength{\tabcolsep}{5pt}
\renewcommand{\arraystretch}{1.5}
\caption{Calculated $E2$ and $M1$ transition amplitudes in a.u..
The corresponding wavelengths ($\lambda$) are presented in \AA}
\vspace{0.5cm}
\begin{tabular}{rcrrrrrrr}
\hline

 & & \multicolumn{3}{|c|}{$E2$} & \multicolumn{3}{|c|}{$M1$} \\
Transitions & $\lambda$ & DF & Corr & RCC & DF & Corr & RCC\\

\hline

$5s$ $^{2}S_{1/2}\rightarrow 5d$ $^{2}D_{3/2}$ & 778.50   & 6.8096 & -0.6241 & 6.1855 &        &       &      \\

                $\rightarrow 5d$ $^{2}D_{5/2}$ & 776.74   & 8.3060 & -0.7744 & 7.5316 &        &       &      \\


$5p$ $^{2}P_{1/2}\rightarrow 5p$ $^{2}P_{3/2}$ & 23030.33 & 8.7362 & -0.7523 & 7.9839 &1.1532  &0.0001 &1.1533\\

                $\rightarrow 6p$ $^{2}P_{3/2}$ & 1126.87  & 5.1577 & -0.4127 & 4.7450 &0.0317  &0.0011 &0.0328\\

                $\rightarrow 4f$ $^{2}F_{5/2}$ & 954.31   & 11.5617& -1.0246 & 10.5371&        &       &      \\

$5p$ $^{2}P_{3/2}\rightarrow 6p$ $^{2}P_{1/2}$ & 1203.94  & 6.1114 & -0.4543 & 5.6571 &0.0336  &0.0008 &0.0344\\

                $\rightarrow 6p$ $^{2}P_{3/2}$ & 1184.84  & 5.7550 & -0.4327 & 5.3223 &0.0003  &0.0006 &0.0009\\

                $\rightarrow 4f$ $^{2}F_{5/2}$ & 995.56   & 6.5714 & -0.5601 & 6.0113 &        &       &      \\

                $\rightarrow 4f$ $^{2}F_{7/2}$ & 995.48   & 16.1049& -1.3733 & 14.7316&        &       &      \\

$6s$ $^{2}S_{1/2}\rightarrow 5d$ $^{2}D_{3/2}$ & 63347.27 & 20.8068& -1.1669 & 19.6399&        &       &      \\

                $\rightarrow 5d$ $^{2}D_{5/2}$ & 53524.59 & 25.7109& -1.4279 & 24.2830&        &       &      \\


$5d$ $^{2}D_{3/2}\rightarrow 5d$ $^{2}D_{5/2}$ & 345184.67& 12.5903& -0.8226 & 11.7677&1.5491  &0.0001 &1.5492\\

                $\rightarrow 7s$ $^{2}S_{1/2}$ & 2440.39  & 6.9372 & -0.3581 & 6.5791 &        &       &      \\

$5d$ $^{2}D_{5/2}\rightarrow 7s$ $^{2}S_{1/2}$ & 2457.77  & 8.7467 & -0.4355 & 8.3112 &        &       &      \\

$6p$ $^{2}P_{1/2}\rightarrow 6p$ $^{2}P_{3/2}$ & 74682.60 & 37.2585& -2.0824 & 35.1762&1.1531  &0.0000 &1.1531\\

                $\rightarrow 4f$ $^{2}F_{5/2}$ & 5752.02  & 40.9209& -2.2989 & 38.6220&        &       &      \\

$6p$ $^{2}P_{3/2}\rightarrow 4f$ $^{2}F_{5/2}$ & 6232.01  & 22.1480& -1.2351 & 20.9129&        &       &      \\

                $\rightarrow 4f$ $^{2}F_{7/2}$ & 6228.90  & 54.1958& -3.0186 & 51.1772&        &       &      \\

$4f$ $^{2}F_{5/2}\rightarrow 4f$ $^{2}F_{7/2}$ & 12500000 & 17.8540& -1.0507 & 16.8033& 1.8516&0.0001  &1.8517\\

\hline
\end{tabular}
\label{Table:III}
\end{table*}

\begin{table*}
\centering
\caption{Calculated lifetimes using the relativistic coupled-cluster
(RCC) theory of some low-lying states in $10^{-9}$ sec along with
their comparisons with the other theoretical (Others) and
experimental (Exp) results.}
\begin{tabular}{l}
\begin{tabular}{lrcrc}
\hline
State & RCC & RMBPT$^c$ & Others & Exp \\
\hline
\\
$5p$ $^{2}P_{1/2}$ & 1.78 & 1.84 & $1.26^{a}$ & $1.45 \pm 0.10^{b}$, $1.72\pm 0.07^{d}$\\
$5p$ $^{2}P_{3/2}$ & 1.42 & 1.45 &            & $1.50\pm 0.15^{d}$\\
$6s$ $^{2}S_{1/2}$ & 0.65 &      &            & \\
$5d$ $^{2}D_{3/2}$ & 0.53 & 0.67 & $0.56^{a}$ & $0.75\pm 0.06^{d} $\\
$5d$ $^{2}D_{5/2}$ & 0.58 & 0.61 &            & $0.98\pm 0.10^{b}$, $0.58\pm 0.05^{d}$\\
$6p$ $^{2}P_{1/2}$ & 4.40 &      &            & \\
$6p$ $^{2}P_{3/2}$ & 4.54 &      &            & \\
$4f$ $^{2}F_{5/2}$ & 1.70 & 1.82 &            & $1.70\pm 0.07^{d}$\\
$4f$ $^{2}F_{7/2}$ & 1.72 & 1.74 &            & $1.72\pm 0.07^{d}$\\
$7s$ $^{2}S_{1/2}$ & 1.03 &      &            & \\
$5g$ $^{2}G_{7/2}$ & 2.65 &      &            & \\
$5g$ $^{2}G_{9/2}$ & 2.66 & 2.71 &            & $2.84\pm 0.30^{d}$\\
\\
\hline
\\
\end{tabular}
\\
$a\rightarrow$ Relativistic Hartree Fock (RHF): \cite{cheng};\\
$b\rightarrow$ Beam Foil: \cite{TA};\\
$c\rightarrow$ RMBPT: \cite{SV};\\
$d\rightarrow$ Beam Foil: \cite{WA};
\end{tabular}
\label{Table:IV}
\end{table*}

\begin{table*}
\centering
\caption{Calculated hyperfine $A$ constants along with the
different correlation contributing terms in MHz.}
\begin{tabular}{c r r r r r r r r r}
\hline\hline
State \hspace{0.3cm} & $d$ \hspace{0.3cm} & $\overline{d}$ \hspace{0.3cm} & $\overline{d}S_{1}$ \hspace{0.3cm} & $\overline{d}S_{2}$ \hspace{0.3cm} & $S_{1}\overline{d}S_{1}$ \hspace{0.3cm}  & $S_{1}\overline{d}S_{2}$ \hspace{0.3cm} & $S_{2}\overline{d}S_{2}$ \hspace{0.3cm}  & Norm. \hspace{0.3cm} & RCC\\
\\
\hline
\\
$5s$ $^{2}S_{1/2}$ &17672.45 & 17635.18 & 2392.83  & 1382.51 & 81.21 & 72.73 & 345.29  & -395.71 & 21358.30\\
$5p$ $^{2}P_{1/2}$ &3317.77  & 3315.07  & 576.20  & 218.52  & 25.13 & 17.50 & 48.24   & -66.76  & 4107.07 \\
$5p$ $^{2}P_{3/2}$ &514.01   & 518.29   & 89.92   & 68.51   & 3.92  & 5.10  & 22.45   & -10.91  & 693.70  \\
$6s$ $^{2}S_{1/2}$ &4792.83  & 4779.14  & 318.36  & 340.26  & 5.32  & 2.40  & 106.62  & -47.66  & 5467.19 \\
$5d$ $^{2}D_{3/2}$ &97.59    & 100.98   & 19.73   & 22.30   & 0.99  & 1.40  & 5.26    & -1.16   & 149.53  \\
$5d$ $^{2}D_{5/2}$ &41.01    & 42.38    & 8.24    & 11.11   & 0.41  & 0.66  & 0.33    & -0.49   & 62.63   \\
$6p$ $^{2}P_{1/2}$ &1090.79  & 1089.88  & 97.20   & 76.63   & 2.22  & 2.27  & 12.99   & -11.97  & 1262.70 \\
$6p$ $^{2}P_{3/2}$ &173.61   & 174.80   & 16.01   & 23.18   & 0.38  & 0.53  & 11.54   & -2.11   & 223.49  \\
$4f$ $^{2}F_{5/2}$ &2.60     & 2.65     & 0.48    & -2.52   & 0.03  & -0.24 & 1.73    & -0.01   & 2.14    \\
$4f$ $^{2}F_{7/2}$ &1.45     & 1.49     & 0.27    & -3.32   & 0.02  & -0.41 & 0.03    & 0.01    & -1.91   \\
$7s$ $^{2}S_{1/2}$ &2234.71  & 2227.97  & -0.49   & 153.23  & 0.01  & -4.34 & 53.48   & -16.63  & 2396.80 \\
\\
\hline\hline
\end{tabular}
\label{Table:V}
\end{table*}

\begin{table*}
\caption{Calculated hyperfine $B$ constants along with the
different correlation contributing terms in MHz.}
\begin{tabular}{c r r r r r r r r r}
\hline\hline
State \hspace{0.3cm} & $d$ \hspace{0.3cm} & $\overline{d}$ \hspace{0.3cm} & $\overline{d}S_{1}$ \hspace{0.3cm} & $\overline{d}S_{2}$\hspace{0.3cm} & $S_{1}\overline{d}S_{1}$ \hspace{0.3cm} & $S_{1}\overline{d}S_{2}$ \hspace{0.3cm} & $S_{2}\overline{d}S_{2}$ \hspace{0.3cm} & Norm. \hspace{0.3cm} & RCC\\
\\
\hline

$5p$ $^{2}P_{3/2}$& 648.57 & 651.06 & 113.13 & 138.88 & 4.94 & 6.17 & 10.05 & -14.25 & 905.47\\
$5d$ $^{2}D_{3/2}$& 41.02  & 42.56  & 8.32   & 45.33  & 0.42 & 1.76 & -0.06 & -0.76  & 97.57\\
$5d$ $^{2}D_{5/2}$& 55.79  & 57.76  & 11.22  & 63.38  & 0.56 & 2.43 & -0.30 & -1.04  & 134.00\\
$6p$ $^{2}P_{3/2}$& 219.07 & 219.75 & 20.19  & 41.75  & 0.47 & 0.37 & 3.51  & -2.66  & 282.32\\
$4f$ $^{2}F_{5/2}$& 1.76   & 1.86   & 0.34   & 28.28  & 0.03 & 1.30 & -0.30 & -0.15  & 31.36\\
$4f$ $^{2}F_{7/2}$& 2.06   & 2.17   & 0.40   & 33.15  & 0.03 & 1.53 & -0.31 & -0.18  & 36.79\\
$5g$ $^{2}G_{7/2}$& 0.31   & 0.31   & 0.00   & 7.08   & 0.00 & 0.04 & -0.05 & 0.00   &7.39\\
$5g$ $^{2}G_{9/2}$& 0.33   & 0.33   & 0.00   & 7.73   & 0.00 & 0.05 & -0.05 & 0.00   &8.06\\

\\
\hline
\end{tabular}
\label{Table:VI}
\end{table*}

\begin{table*}

\centering

\caption{Hyperfine splitting (Spl) of the ground and few low lying
excited states in MHz. The percentage contribution from the $B$
constant (B-Cont) to these splitting are presented at the last
column.}
\begin{tabular}{c}
\begin{tabular}{l c r r}
\hline\hline
State & $F_{1}\leftrightarrow F_{2} $&  Spl & \hspace{0.3cm}B-Cont \\
\hline

\\

$5s$ $^{2}S_{1/2} $ & $5 \leftrightarrow 4$ & 106791.51 &0      \\

$5p$ $^{2}P_{1/2} $ & $5 \leftrightarrow 4$ &  20535.37 &0      \\

$5p$ $^{2}P_{3/2} $ & $6 \leftrightarrow 5$ &   4765.84 & 12.67 \\

                    & $5 \leftrightarrow 4$ &   3279.86 & -5.75 \\

$6s$ $^{2}S_{1/2} $ & $5 \leftrightarrow 4$ &  27335.95 & 0     \\

$5d$ $^{2}D_{3/2} $ & $6 \leftrightarrow 5$ &    962.22 & 6.76  \\

                    & $5 \leftrightarrow 4$ &    727.31 & -2.79 \\

$5d$ $^{2}D_{5/2} $ & $7 \leftrightarrow 6$ &    500.92 & 12.48 \\

                    & $6 \leftrightarrow 5$ &    385.81 & 2.61  \\

$6p$ $^{2}P_{1/2} $ & $5 \leftrightarrow 4$ &   6313.48 & 0     \\

$6p$ $^{2}P_{3/2} $ & $6 \leftrightarrow 5$ &   1529.17 & 12.31 \\

                    & $5 \leftrightarrow 4$ &   1058.64 & -5.56 \\

$7s$ $^{2}S_{1/2} $ & $5 \leftrightarrow 4$ &  11984.02 & 0     \\

\\

\hline

\end{tabular}

\end{tabular}
\label{Table:VII}
\end{table*}

\begin{table*}
\centering 
\caption{Hyperfine anomaly of In III.}
\begin{tabular}{c}
\begin{tabular}{l c c c r r}
\hline\hline
State & A (113)  &  A (115) & A (117)& $^{113}\Delta^{115}(\%)$&$^{115}\Delta^{117}(\%)$ \\
\hline
\\ 
$5s^{2}S_{\frac{1}{2}}$&21313.0636 & 21358.3017 & 21271.3821&0.007491126&0.009042670\\
$6s^{2}S_{\frac{1}{2}}$&5455.6089 & 5467.1903 & 5444.9430&0.007461625&0.009008028\\
$7s^{2}S_{\frac{1}{2}} $&2391.7256 & 2396.8034 & 2387.0509&0.007438228&0.008981253\\
$5p^{2}P_{\frac{1}{2}}$&4098.0919 & 4107.0750 & 4090.7031&0.000558990&0.000676622\\
$5p^{2}P_{\frac{3}{2}}$&692.1824 & 693.6997 & 690.9345&0.000553917&0.000670590\\
$6p^{2}P_{\frac{1}{2}} $&1259.9349 & 1262.6970 & 1257.6639&0.000534891&0.000641958\\
$6p^{2}P_{\frac{3}{2}} $&223.0038 & 223.4925 & 222.6013&0.000642552&0.000779395\\

\hline
\end{tabular}
\end{tabular}
\label{Table:VIII}
\end{table*}

\subsection{Conclusion}
The electro-magnetic transition amplitudes, lifetimes and hyperfine constants
are calculated using a highly correlated theoretical approach with
a proper account of relativity. Our calculated transition
line parameters can be applied for abundance estimations in
different astronomical systems and laboratory plasmas. The ground
state hyperfine splitting of this ion predicts its use as
microwave frequency standard at 10$^{-11}$ sec.
 The detail analysis of the different correlation contributing terms
associated with the coupled-cluster theory show their impacts in
the calculations of the hyperfine constants. The hyperfine $B$
constants of the fine structures of $4f$ $^{2}F $ and $5g$ $^{2}G$ terms are found to be abnormally correlated due to the very strong
influence of the core polarization. The calculated hyperfine
splitting can be used for accurate line-profile analysis of astrophysically important transition lines. We have also observed distinct feature in the hyperfine anomaly parameters compare to neutral Indiun.

\begin{acknowledgments}
We are thankful to Prof. B. P. Das and Dr. R. K. Choudhuri, IIA,
Bangalore, India and Dr. B. K. Sahoo, PRL, Ahmedabad, India for
providing the COUPLED-CLUSTER code to us. We also want to
acknowledge Board of Research in Nuclear Sciences (BRNS), India;
for funding.
\end{acknowledgments}


\begin{thebibliography}{14}
\clearpage
\bibitem{schauer}
M. M. Schauer, J. R. Danielson, D. Feldbaum, M. S. Rahaman, L. B. Wang, J. Zhang, X. Zhao, and J. R. Torgerson, Phys. Rev. A {\bf 82}, 062518 (2010).
\bibitem{jelenkovic}
B. M. Jelenkovic, S. Chang, J. D. Prestage, and L. Maleki, Phys. Rev. A {\bf 74}, 022505 (2006).
\bibitem{dixit}
G. Dixit, H. S. Nataraj, B. K. Sahoo, R. K. Chaudhuri, and S. Majumder, Phys. Rev. A {\bf 77}, 012718 (2008).
\bibitem{majumder}
M. Gunawardena, H. Cao, P. W. Hess, and P. K. Majumder, Phys Rev. A {\bf 80}, 032519 (2009).
\bibitem{eck}
T G Eck, A Lurio, P Kusch, Phys. Rev, {\bf 106} 954 (1957).
\bibitem{Mann}
A. K. Mann And P. Kusch, phys. rev, {\bf 71}, 4 (1950).
\bibitem{Chaudhuri}
M. Das, R. K. Chaudhuri, S. Chattopadhyay and U. S. Mahapatra,J. Phys. B: At. Mol. Opt. Phys. {\bf 44},065003 (2011).
\bibitem{safronovaano}
U. I. Safronova, M. S. Safronova, M. G. Kozlov, Phys Rev. A {\bf 76}, 022501 (2007).
\bibitem{raghavan}
P. Raghavan, At. Data. Nucl. Data Tables {\bf 42}, 189 (1989).
\bibitem{Fortson}
E. N. Fortson, Y. Pang and L. Wilets, Physical rev. Letter {\bf 65}, 23 (1990).
\bibitem{Marie}
A. Marie, M. Pendrill, Hyperﬁne Interactions {\bf 127} 41 (2000). 
\bibitem{gustavsson}
J R C Lopez-Urrutia,P. Beiersdorfer, K. Widmann, B. B. Birkett, A.-M. Martensson-Pendrill and M. G. H. Gustavsson, Phys. Rev. A, {\bf 57}, 879 (1998).
\bibitem{niesen}
S H Devare, H. G. Devare, F Pleiter, F C Magendans and L Niesen, Hyperfine Interactions, {\bf 15/16}, 31 (1983). 
\bibitem{persson}
J R Persson, Atomic Data and Nuclear Data tables, {\bf 99}, 62 (2013).
\bibitem{LUTZ} 
O. Lutz, A. Nolle, and A. Uhl, Z. Physik 248, 159 (1971).
\bibitem{BW}
A. Bohr, V.F. Weisskopf, Phys. Rev. {\bf 77}, 94 (1950).
\bibitem{butt}
S. Buettgenbach, Hyperfine Int, {\bf 20}, 1 (1984).
\bibitem{stroke}
H.H. Stroke, H.T. Duong  and J. Pinard, Hyperfine Int., {\bf 129}, 319 (2000) 
\bibitem{TF}
S. Djeniže, A. Srećković, S. Bukvić, Spectrochimica Acta Part B, {\bf 61}, 588 (2006).
\bibitem{ZM}
Z. Simic, M. S. Dimitrijevic, A. Kovacevic and S. Sahal-Brechot, Baltic Astronomy, {\bf 20}, 613 (2011).
\bibitem{CR}
Cowley C R, Hartoog M R and Cowley A P, Astrophys. J. {\bf 194}, 343 (1974).
\bibitem{vitas}
N. Vitas, I. Vince, M. Lugaro, O. Andriyenko, M. Gosic and R. J. Rutten, Mon. Not. R. Astron. Soc. {\bf 384}, 370 (2008).
\bibitem{LI}
M. Skocic, M. Burger, S. Bukvic and S. Djenize, J. Phys. B: At. Mol. Opt. Phys. {\bf 45}, 225701 (2012).
\bibitem{DJ}
Djenize S, Sreckovic A and Bukvic S,  Spectrochim. Acta B {\bf 61}, 588 (2006).
\bibitem{Dimitrijevic}
Dimitrijevic M, Kovacevic A, Simic Z and Sahal-Brechot S Balt. Astron.
{\bf 20} 495 (2011).
\bibitem{ali}
M. A. Ali, and Yong-Ki Kim, Phys. Rev. A {\bf 38}, 3992 (1988).
\bibitem{ND}
Nodwell R A, PhD Thesis University of British Columbia, Vancouver (1956).
\bibitem{SK}
 K S Bhatia, J. Phys. B: Atom. Molec. Phys., {\bf 11}, 14 (1978).
 \bibitem{TA}
T. Andersen, A. Kirkegard Nielsen and G. Sorensen, Physica Scripta. {\bf 6}, 122 (1972).
\bibitem{SV}
U. I. Safronova, I. M. Savukov, M. S. Safronova, W. R. Johnson, Physical Review A {\bf 68}, 062505 (2003).
\bibitem{GLO}
L. Głowacki and J. Migdałek, Phys. Rev. A {\bf 80}, 042505  (2009).
\bibitem{cheng}
Kwok-tsang Cheng and Yong-Ki Kim,J. Opt. Soc. Am., {\bf 69}, 1 (1979).
\bibitem{martin}
I. Martin, M. A. Almaraz, C. Lavin,Z. Phys. D {\bf 35}, 239 (1995).
\bibitem{XB}
Xiao-Bin Ding, F. Koike, I. Murakami, D. Kato,H. A Sakaue, Chen-Zhong Dong and N. Nakamura J. Phys. B: At. Mol. Opt. Phys. {\bf 45}, 035003 (2012).
\bibitem{dutta}
N. N. Dutta, and S. Majumder, Phys. Rev. A {\bf 85}, 032512 (2012).
\bibitem{Dutta}
N. N. Dutta, and S. Majumder, Astrophys. J. {\bf 737}, 25 (2011).
\bibitem{Bishop}
R. F. Bishop, and H. G. KŁummel, Phys. Today, March {\bf 40}, 52 (1987).
\bibitem{Lindgren}
I. Lindgren, and D. Mukherjee, Phys. Rep 151, 93 (1987).
\bibitem{Sahoo}
R. K. Chaudhuri, B. K. Sahoo, B. P. Das, H. Merlitz, U.
S. Mahapatra, and D. Mukherjee, J. Chem. Phys. 119,
10633 (2003).
\bibitem{szabo}
A. Szabo and N. S. Ostlund, {\it Modern Quantum Chemistry:
Introduction to Advanced Electronic Structure Theory} (Dover
Publications, Mineola, 1996).
\bibitem{sourav}
N. N. Dutta,S. Roy, G. Dixit, S. Majumder, Phys Rev. A {\bf 87}, 012501 (2013).
\bibitem{Lutz}
O. Lutz, A. Nolle, and A. Uhl, Z. Physik {\bf 248}, 159 (1971).
\bibitem{Dutta2013}
N. N. Dutta, and S. Majumder, Phys. Rev. A {\bf 88}, 062507 (2013)
\bibitem{NI}
NIST Atomic Spectra Database (ver.5.0), [Online]. Available: http://physics.nist.gov/asd [2012, August 1]. National Institute of Standards and Technology, Gaithersburg, MD.
\bibitem{JM}
J Migdalek† and M Garmulewicz,J. Phys. B: At. Mol. Opt. Phys. {\bf 33}, 1735 (2000).
\bibitem{WA}
W. Ansbacher, E.H.Pinnington, J. A. Kernahan, R. N. Gosselin, Can. J. Phys. {\bf 64}, 1365 (1986).
\bibitem{gopal}
G Dixit, B. K. Sahoo, R. K. Chaudhuri and S. Majumder, Phys. Rev. A {\bf 76}, 042505 (2007).
\bibitem{Sancho}
P. Sancho, Arxiv:1311.0998 (2013).
\bibitem{büttgenbach}
S. Büttgenbach, Hyperfine Int. {\bf 20} 1 (1984).
\bibitem{Owusu}
A. Owusu, X. Yuan, S. N. Panigrahy, R. W. Dougherty, and T. P. Das, Phys. Rev. A {\bf 55}, 4 (1997).
\bibitem{chou}
H.S. Chou and W.R. Johnson, Phys. Rev. A {\bf 56}, 2424  (1997).


\end{thebibliography}
\end{document}